\documentstyle[pra,aps,epsfig,amsfonts,amsmath]{revtex}

\draft

\begin{document}

\title{Momentum transferred to  a trapped Bose-Einstein
condensate \\  by stimulated light scattering}

\author{A. Brunello$^1$, F. Dalfovo$^2$, L. Pitaevskii$^{1,3}$,
S. Stringari$^1$ and F. Zambelli$^1$}

\address{$^1$ Dipartimento di Fisica, Universit\`a di Trento, and \\
Istituto Nazionale per la Fisica della Materia,
I-38050 Povo, Italy}

\address{$^2$ Dipartimento di Matematica e Fisica, Universit\`a
Cattolica del Sacro Cuore, and \\
Istituto Nazionale per la Fisica della Materia,
gruppo collegato di Brescia, \\
Brescia, Italy }

\address{$^3$ Kapitza Institute for Physical Problems,
ul. Kosygina 2, 117334 Moscow}

\date{\today}

\maketitle

\begin{abstract}
The response of a trapped Bose-Einstein condensed gas to a density
perturbation generated by a two-photon Bragg pulse is investigated by
solving the time-dependent Gross-Pitaevskii equation. We calculate
the total momentum imparted to the condensate as a function of both
the time duration of the pulse and the frequency difference of the
two laser beams. The role of the dynamic response function in
characterizing the time evolution of the system is pointed out, with
special emphasis to the phonon regime. Numerical simulations are
compared with the predictions of local density approximation.
The relevance of our results for the interpretation of current
experiments is also discussed.
\end{abstract}


\section{Introduction}

Recent experiments based on two-photon Bragg spectroscopy \cite{MIT1,MIT2}
have given first valuable information on the density-density response
function of a trapped Bose-Einstein condensate, thereby providing an 
important opportunity to understand the role of quantum correlations and to 
check theoretical predictions on the dynamics of nonuniform condensates at 
both low and high excitation energy \cite{zambelli,brunello,NZ}. In these 
experiments two laser beams impinge upon a condensate, whose 
atoms can undergo stimulated light scattering events by absorbing a
photon from one of the beams and emitting into the other. The difference
in the wave vectors of the beams defines the momentum transfer in a single
scattering event,  $\hbar {\mathbf q}$ , while the frequency difference
defines the corresponding energy transfer, $\hbar \omega $. Both the values 
of ${\mathbf q}$ and $\omega $ can be tuned by changing the angle between the
two beams and varying their frequency difference.  For small values of 
${\mathbf q}$ the system is excited in the phonon regime and the response 
is usefully detected by measuring the net momentum imparted to the gas. 
The latter is measured by observing the center of mass motion of 
the sample after releasing the trap. 

Since the momentum acquired by the condensate is the physical quantity 
measured in these experiments,  it is important to understand what are the 
underlying mechanisms characterizing its behavior and what can be 
learnt from a careful study of its time dependence. In this work we 
investigate this problem by using the Gross-Pitaevskii (GP) theory
for the order parameter of the condensate, with special emphasis on 
the phonon (low $q$) regime. First we present results of simulations based 
on the numerical integration of the time dependent GP equation. Then we 
analyze the same results from the viewpoint of linear response theory, 
testing the accuracy of the local density approximation for the dynamic 
structure factor. Finally we compare our results with the experimental 
data obtained from two-photon Bragg spectroscopy \cite{MIT2}.

\section{Time dependent Gross-Pitaevskii equation}

The interaction of the atoms with the laser field is described
by the Hamiltonian 
\begin{equation}
H_{\rm Bragg} = {\frac{V}{2}}
\left( \delta \rho _{{\mathbf q}}^{\dagger }e^{-i\omega
t}+\delta \rho _{{\mathbf q}}e^{i\omega t}\right)
\label{eq:Hbragg}
\end{equation}
where
$\delta\rho_{{\mathbf q}}=\rho_{{\mathbf q}}-\langle\rho_{{\mathbf q}}
\rangle$
is the fluctuation of the density operator
\begin{equation}
\rho _{{\mathbf q}}=\sum_{j=1}^{N}e^{-i{\mathbf q\cdot r}_{j}}\;,
\label{eq:rhoq}
\end{equation}
while the strength of the perturbation, $V$, is proportional to the 
intensity of  the laser beam and can be related to the two-photon 
Rabi frequency \cite{Hbragg}.
The validity of expression (\ref{eq:Hbragg}) is not limited to the
regime of small perturbations, but can also be used to explore
non linear effects, including, for example, the coherent splitting of the
condensate \cite{phillips}.

We simulate the time evolution of the system by solving the
Gross-Pitaevskii (GP) equation
\begin{equation}
i\hbar\frac{\partial}{\partial t} \Phi ({\mathbf r},t) =
\left[-\frac{\hbar^2}{2m}\nabla^2 + V_{\rm ho}({{\mathbf r}})
+ g|\Phi|^2+\vartheta (t) V \cos (qz - \omega t) 
 \right]\Phi ({\mathbf r},t)
\label{eq:GP}
\end{equation}
for the order parameter of the gas, $\Phi({\mathbf r},t)$. The harmonic
confinement is represented by the potential
$V_{\rm ho} ({{\mathbf r}}) = (m/2) [ \omega_\perp^2 r_\perp^2 +
\omega_z^2 z^2]$, with $r_{\perp}^2 = x^2 + y^2$, while the coupling
constant in the mean field term, $g=4\pi \hbar^2 a/m$, is fixed by the
$s$-wave scattering length $a$. The last term in the r.h.s. comes from
Eq.~(\ref{eq:Hbragg}), while $\vartheta(t)$ is the Heaviside step function.
The GP equation (\ref{eq:GP}) is expected to provide an accurate description
of the system at sufficiently low temperature \cite{RMP}. We numerically
solve it for a system of sodium atoms confined by a cigar-shaped potential,
with $\omega_{\perp}= 8 \omega_z$. For the Thomas-Fermi parameter 
$15Na/a_{\perp}$ we choose the value $10^4$. Here $a_{\perp}= 
\sqrt{\hbar /m\omega_{\perp}}$ is the oscillator length in the 
transverse direction. Actually, by a proper rescaling of the spatial 
coordinates (in units of $a_{\perp}$) and energies (in units of 
$\hbar \omega_{\perp}$), the solutions of the time-dependent 
GP equation (\ref{eq:GP}) depend only on the Thomas-Fermi parameter, 
the ratio $\omega_z/\omega_{\perp}$ and, of course, the parameters 
characterizing the Bragg potential. Finally, we  take
${\mathbf q}$ along $z$ to  preserve the axial symmetry of the
problem as in the MIT experiment \cite{MIT2}. In order to select the
phonon regime, one has to choose a wave vector $q$ such that $q\xi < 1$,
where $\xi$ is the healing length. The latter is defined as $\xi = [8\pi n
a]^{-1/2}$, where $n$ is the density evaluated in the center of the
trap. In our calculation, a typical value is $q=3 a_{\perp}^{-1}$, with
$q\xi =0.42$.

The numerical integration of the time dependent GP equation (\ref{eq:GP})
is performed by using a method previously developed in \cite{modugno},
which is suitable for axially symmetric condensates. The equation is mapped
on a two-dimensional grid of points and solved by using a combination of
Fast Fourier and Crank-Nicholson integration algorithms.  From the solution
of Eq.~(\ref{eq:GP}), one can easily calculate net momentum imparted to 
the condensate  by direct integration of the current density associated 
with the order parameter
\begin{equation}
P_z(t) = { \hbar \over 2 i} \!\int\!\!d{\mathbf r}\,
\Phi^*({\mathbf r},t)\nabla_z \Phi({\mathbf r},t) \; + \; {\rm c.c.} \; \; .
\label{eq:Pt}
\end{equation}

\begin{center}
\begin{figure}
\input{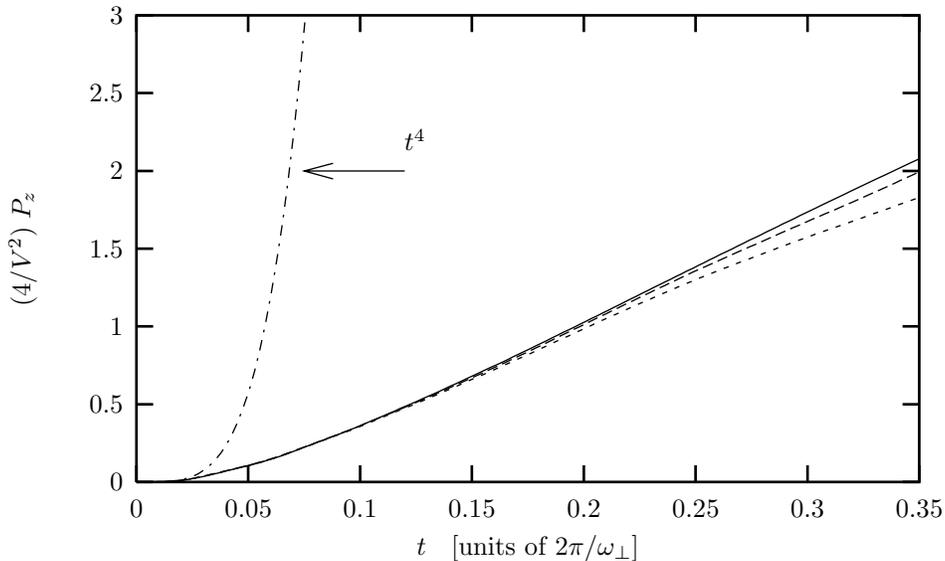}
\caption{
The quantity $(4/V^2) P_z$, in units of $[\hbar \omega_\perp^2 a_\perp
]^{-1}$ is plotted against time, in units of $\nu_\perp^{-1} = 2\pi /
\omega_\perp $. The Bragg pulse is applied starting at $t=0$, with $q=3
a_\perp^{-1}$ and $\omega=13 \omega_\perp$. 
The dot-dashed line is the model independent short time
expansion (\ref{eq:Plowt}). The solid line is the prediction of LDA, obtained
by using the dynamic structure factor (\ref{eq:SLDATF}) in
Eq.~(\ref{eq:dPdt}). The two dashed lines are the results of the numerical
integration of the GP equation (\ref{eq:GP}), with two different values of
the Bragg intensity $V$, namely, $V=0.2 \hbar \omega_\perp$ (long dashed) 
and $0.4 \hbar \omega_\perp$ (short-dashed). }
\label{fig1}
\end{figure}
\end{center}

In Fig.~\ref{fig1} we show the results obtained for $P_z(t)$,
having fixed the frequency difference of the Bragg pulses to
be $\omega =13 \omega_\perp$. We use different values of the strength $V$ 
in order to check whether we are in the appropriate limit of linear 
response to the external perturbation. In the linear regime the momentum 
$P_z$ of the condensate should depend quadratically on the intensity $V$ 
of the Bragg pulse. We have verified that, by taking $V$ smaller than 
$0.2 \hbar \omega_\perp$, the linear regime is ensured  for times shorter 
than $0.2 \, (2\pi/\omega_\perp)$.  We have solved the time dependent GP 
equation  also in the case of a strong Bragg pulse ($V > 10 \hbar 
\omega_\perp$).  For such a pulse the change in the momentum
distribution of the condensate is huge \cite{brunello} and
our simulation shows that, in this case, the  momentum transfer stays
systematically below the value predicted by linear theory.

In Fig.~\ref{fig2} the net momentum transfer is shown as a function
of the frequency $\omega$ for two different choices of the time duration
$\tau_{B}$ of the Bragg pulse. The figure shows that the shape of the signal
exhibits a significant time dependence, becoming narrower by increasing
$\tau_{B}$.

\section{Linear response theory}

In order to better understand the  behavior of the momentum transfer
as a function of  $t$ and $\omega$, we use the formalism
of linear  response function, valid for weak intensities of the
Bragg pulse. 

First, we write the Heisenberg equation of motion for the momentum 
$P_{z}= \langle \sum_{j=1}^{N} p_{j}^{z}\rangle $:
\begin{equation}
\frac{dP_{z}(t)}{dt} 
=\frac{1}{i\hbar}\langle [\sum_{j=1}^{N} p_{j}^{z} , H_{\rm tot}] 
\rangle
\label{eq:heisenberg}
\end{equation}
where the Hamiltonian of the system is 
\begin{equation}
H_{\rm tot}= \sum_{i=1}^N {p_i^2 \over 2m} + \sum_{i=1}^N
V_{\rm ext}({\mathbf r}_i)+g\sum_{i<j}\delta({\mathbf r}_i-{\mathbf
r}_j)+\vartheta(t)H_{\rm Bragg} \; .
\label{eq:Htot}
\end{equation}  
The commutator can be explicitly evaluated, obtaining the exact 
equation
\begin{equation}
\frac{dP_{z}(t)}{dt} = -m \omega_z^2 Z - \frac{iqV}{2}
\left( \langle \delta \rho _{{\mathbf q}}^{\dagger }\rangle
e^{-i\omega t}-\langle \delta 
\rho _{{\mathbf q}}\rangle e^{i\omega t} \right) \; . 
\label{eq:PH}
\end{equation}
The first term on the r.h.s  originates from the confining oscillator 
potential, while the second one from the Bragg perturbation. The 
quantity $Z= \langle \sum_{j=1}^{N}z_j\rangle$ is the expectation 
value of the $z$-th component of the center of mass coordinate. 
Notice that the commutator in Eq.~(\ref{eq:heisenberg}) 
is not affected by the kinetic energy term, nor by two-body
interactions which are translational invariant quantities.

Now, one can use the linear response theory for the density fluctuations.  
A convenient way consists in writing the time dependent perturbation as
\begin{equation}
\vartheta(t)H_{\rm Bragg} = \frac{iV}{4\pi}\delta
\rho _{{\mathbf q}}^{\dagger} \int d\omega^{\prime }\ 
\frac{e^{-i\omega ^{\prime }t}}{\omega^{\prime
}-\omega +i\eta } \, + \, H.c. \; \; ,
\label{eq:Hbraggtheta}
\end{equation}
which follows from the Fourier representation of the step function. 
This perturbation induces density fluctuations in the form \cite{PN}
\begin{equation}
\langle\delta\rho_{{\mathbf q}}\rangle = 
- \frac{V e^{-i\omega t}}{2\hbar}
 \int\!d\omega ^{\prime} \left[S({\mathbf q},\omega^{\prime})
-S(-{\mathbf q},-\omega^{\prime}) \right]   
{\frac{e^{i(\omega - \omega ^{\prime}) t}-1}
{\omega - \omega ^{\prime }}}\;,
\label{eq:deltarhoq}
\end{equation}
where
\begin{equation}
S({\mathbf q},\omega )={\frac{1}{\cal Z}}
\sum_{m,n}e^{-\beta E_{m}}\mid \langle m\mid \delta
\rho _{{\mathbf q}}\mid n \rangle \mid ^{2}\delta (\omega -\omega _{mn})
\label{eq:S}
\end{equation}
is the dynamic structure factor of the system, with $\beta=1/kT$,
$\hbar \omega_{mn}= E_m - E_n$ and ${\cal Z}$ is the usual
partition function. In deriving this result, one assumes that the 
fluctuation $\langle\delta\rho_{{\mathbf q}}\rangle$ induced by the
$e^{-i\omega^{\prime}t}$ component of the perturbation
(\ref{eq:Hbraggtheta}) oscillates like $e^{-i\omega^{\prime}t}$, so
neglecting the term oscillating like $e^{i\omega^{\prime}t}$. The latter
exactly vanishes in a uniform body and, in general,  is exponentially
small if the value of $q$ is much larger than the inverse of the size
of the system. In the following we will always neglect this contribution.

Inserting the density fluctuation (\ref{eq:deltarhoq}) into 
Eq.~(\ref{eq:PH}), one gets the following equation for the rate of 
momentum transfer at $t>0$:
\begin{equation}
{\frac{dP_z(t)}{dt}}=-m\omega^2_z Z+{2 q \over \hbar}
\left({\frac{V}{2}}\right)^2
\int d\omega ^{\prime }
\left[S({\mathbf q},\omega^{\prime })-
S(-{\mathbf q},-\omega^{\prime })\right]
{\sin\left[(\omega -\omega^{\prime}) t\right]
\over \omega -\omega^{\prime}} \; .
\label{eq:dPdt}
\end{equation}
This equation explicitly contains the effect of the confining potential 
through the first term in the r.h.s.; this term is small for times  
short compared to the oscillator period $2\pi/\omega_{z}$. We have 
verified that for the times considered in our simulation the effect of 
the external potential in the equation for the momentum rate can be
safely ignored.

For very short times the  expansion of (\ref{eq:dPdt}) yields the model
independent result
\begin{equation}
\frac{dP_z(t)}{dt} = \frac{\omega V^2q^3t^3}{3m}
\label{eq:Plowt}
\end{equation}
where we have made use of both the $f$-sum rule  $\int d\omega \, \omega
S({\mathbf q},\omega)= N\hbar q^2/2m$ and the  relationships 
\begin{equation}
\int\!d\omega \left[ S({\mathbf q},\omega)-
S(-{\mathbf q},-\omega)\right] =
\int d\omega \, \omega^2 \left[S({\mathbf q},\omega)
-S(- {\mathbf q},-\omega)\right]=0 \; . 
\end{equation}

A  closed equation for $P_z(t)$  can be obtained by taking the time
derivative of (\ref{eq:dPdt}) and using the exact equation
$dZ(t)/dt = P_z(t)/m$.   If the duration  of the pulse is
short compared to the oscillator time, but at the same time
large compared to the inverse of the frequency of the applied field,
then the equation for the momentum  rate approaches the golden
rule result:
\begin{equation}
{\frac{dP_z(t)}{dt}}= q \left(\frac{V}{2}\right)^2\frac{2\pi}{\hbar}
\left[ S({\mathbf q},\omega )-S(-{\mathbf q},-\omega )\right]\;.
\label{eq:equiv}
\end{equation}
Although restrictive, the conditions $\omega t\gg 1$ and $\omega _{z}t
\ll 1$, needed to derive result (\ref{eq:equiv}), are  compatible. 
In principle,  even in the
phonon regime where the excitation energy $\hbar \omega $ should be
smaller than the chemical potential $\mu$, the two conditions can be
simultaneously satisfied if one works in the Thomas-Fermi regime where
$\mu \gg \hbar \omega _\perp$. In practice,
we have found difficulties in exploring the large time domain because
of the occurrence of significant nonlinear effects.

Analogous results, valid in the linear regime, can be obtained for
the energy rate. In this case one finds the standard result of
perturbation theory,
\begin{equation}
{\frac{dE(t)}{dt}}=\frac{V^2}{2\hbar}
\int d\omega^{\prime }\, \omega ^{\prime }
\left[ S({\mathbf q},\omega^{\prime } )-S(-{\mathbf q},-\omega^{\prime })
\right] {\sin\left[(\omega- \omega ^{\prime}) t\right]
\over \omega - \omega ^{\prime}} \; ,
\label{eq:dEdt}
\end{equation}
which yields the golden rule result
\begin{equation}
\frac{dE(t)}{dt} = \omega \left(\frac{V}{2}\right)^2\frac{2\pi}{\hbar}
 \left[ S({\mathbf q},\omega )- S(-{\mathbf q},-\omega)\right] 
\label{eq:golden}
\end{equation}
in the large $t$ limit. From the experimental viewpoint it is very
difficult to extract the net energy transfer since it would require a
very high precision in the determination of the release energy.
Furthermore, the release energy does not coincide with the  total energy
of the sample which includes also the  contribution of the confining
potential.

It is   worth noticing that  result (\ref{eq:dPdt}) for the
momentum rate, like Eq.~(\ref{eq:dEdt}) for the energy rate,
is sensitive  to the difference $[ S({\mathbf q},\omega )-
S(-{\mathbf q},-\omega)]$ rather than to the dynamic structure factor
itself. This difference characterizes the imaginary part of the
density-density response function
\begin{equation}
{\rm Im} \left( \chi_{{\mathbf q}} (\omega) \right) = -(\pi/\hbar)
\left[ S({\mathbf q},\omega )-S(-{\mathbf q},-\omega)\right]
\label{eq:chi}
\end{equation}
and follows from the fact that atoms can scatter by absorbing a photon
from either of the laser beams. This represents  an important difference
with respect to other scattering experiments (like, for example, neutron 
scattering from helium) where, by detecting the scattered probe, one 
instead measures directly the dynamic structure factor. The dynamic 
structure factor and the imaginary part of $\chi_{{\mathbf
q}}(\omega)$ have the same behavior 
at $T=0$ since in this case $S({\mathbf q},\omega )$ is zero for negative 
$\omega $. They instead differ at finite temperatures if $k_{B}T$ is of 
the order or higher than the excitation energy $\hbar \omega $. Actually 
the difference (\ref{eq:chi}) significantly suppresses the thermal effects 
exhibited by the dynamic structure factor so that, by measuring 
$\chi_{{\mathbf q}}(\omega)$, one has an easy access to the zero 
temperature value of $S({\mathbf q},\omega )$. For this
reason, even if experiments are carried out at temperatures which do not
satisfy the condition $kT \ll \hbar \omega$, one can safely restrict the
theoretical analysis to the simpler $T=0$ case, provided the temperature
is sufficiently low to ignore the effects due to the thermal depletion
of the condensate.

\section{Local Density Approximation}

If the value of the wave vector $q$  is larger than the inverse of the
size of the condensate the response of the system can be obtained, in 
first approximation, using a local density approximation (LDA). In fact, 
in this case the effects of discretization in the excitation spectrum can
be safely ignored and the system can be treated as a locally uniform
medium. In our simulation we have $qR_{\perp}=21$ and $qR_z=168$ and
consequently this approximation should be well satisfied. In the local
density approximation the dynamic structure factor is given by 
the analytic expression \cite{MIT2,zambelli}
\begin{equation}
S_{\rm LDA}({\mathbf q},\omega )= 
\frac{15\hbar}{8}
\frac{(\hbar ^{2}\omega^{2}-E_{r}^{2})}{E_{r}\mu ^{2}} 
\left[ 1-{\frac{(\hbar ^{2}\omega
^{2}-E_{r}^{2})}{2E_{r}\mu }} \right]^{1/2}
\label{eq:SLDATF}
\end{equation}
 where
\begin{equation}
E_{r}={\frac{\hbar ^{2}q^{2}}{2m}}
\label{eq:Er}
\end{equation}
is the free recoil energy and $\mu = \hbar^2/(2m\xi^2)$ is
the Thomas-Fermi value of the chemical potential. In deriving result
(\ref{eq:SLDATF}) one has to average the Bogoliubov expression
$S({\mathbf q},\omega)= [ \hbar^2q^2/ (2m \epsilon(q))] \delta(\omega -
\epsilon(q)/\hbar)$  over the Thomas-Fermi density profile 
$n({\mathbf r}) = (1/g) [ \mu - V_{\rm ext}({\mathbf r})]$, by 
evaluating the Bogoliubov excitation spectrum
\begin{equation}
\epsilon(q) = \left[ {\hbar^2 q^2 \over 2m}\left({\hbar^2q^2 \over 2m} 
+gn({\mathbf r})\right) \right]^{1/2}
\label{eq:epsilonq}
\end{equation}
at the corresponding density.
In order to apply the LDA,  the momentum transfer $\hbar q$ should not
be however too large, because this approximation ignores the
Doppler effect associated with the spreading of the momentum distribution
of the condensate which is expected to become the leading effect
at very large values of $q$ \cite{zambelli}. This happens for values
of $q$ much larger than the inverse of the healing length, a situation
that is not considered in the present work.

Equation (\ref{eq:SLDATF}) shows that, differently from the case of a
uniform gas, the dynamic structure factor of a trapped condensate
is no longer a delta function, its value being different from zero in
the interval $E_{r}<E<E_{r} [1+2\mu /E_{r}]^{1/2}$. The value $E=E_{r}$
corresponds to the excitation energy in the region near the
border where the gas is extremely dilute and hence noninteracting. The
value $E=E_{r} [1+2\mu /E_{r}]^{1/2}$ is instead the excitation energy of a
Bogoliubov gas evaluated at the central density. Notice that the LDA
expression (\ref{eq:SLDATF}) for $S({\mathbf q},\omega )$ does not depend
on the direction of the vector ${\mathbf q}$ even in the presence of a
deformed trap. 

Results (\ref{eq:dPdt}) and (\ref{eq:SLDATF}) allows one to evaluate 
the time dependence of the momentum imparted to the condensate in 
the linear regime. In Figs.~\ref{fig1} and \ref{fig2}
we compare the results of this local density approximation with the
ones of the numerical integration of the GP equation. The agreement is
excellent and proves that the LDA  is quite adequate to describe 
the response of the system in the conditions considered in 
our analysis. We have also carried out the same comparison for a smaller 
value of momentum transfer ($q=1a_{\perp}^{-1}$) finding a similar good 
agreement.
  
The successful comparison between the predictions of the LDA and the 
numerical simulation  suggests that equations (\ref{eq:SLDATF}) and 
(\ref{eq:dPdt}) are indeed useful for the analysis of the experimental 
results of Ref \cite{MIT2} (see Fig.~\ref{fig3}).
The values of the parameters used in the experiment ($q=4.3 
a^{-1}_{\perp}$, $\omega_z/\omega_{\perp}=0.12$, $q\xi = 0.39$) do 
not differ significantly from the ones used in our simulation, except 
for  the value of the chemical potential which is 
significantly larger ($\mu =60\hbar\omega_{\perp}$). Because of this, 
the numerical simulation of the MIT experiment would be a highly time 
consuming calculation, beyond the purpose of this work. On the other 
hand, in the above experimental conditions the applicability of the
LDA should be even better, because of the increase of the value of 
$qR_{\perp}$ and of $qR_z$. 

In Fig.~\ref{fig3} we compare the LDA prediction (solid line) with 
the available experimental data (points). The latter are given in 
arbitrary units, since the value of $V$ used in \cite{MIT2} is not 
available, and hence the comparison is limited to the shape and 
position of the peak. The LDA curve is obtained by  integrating 
Eq.~(\ref{eq:dPdt}) between $t=0$ and $t=\tau_B$ and using 
expression (\ref{eq:SLDATF}) for the dynamic structure factor. The 
agreement is reasonably good. It is also interesting to compare the 
LDA prediction with the golden rule (\ref{eq:equiv}) integrated over 
the same time interval (dashed line). The LDA curve is broadened 
and quenched as a result of the finite duration of the Bragg pulse,
which enters Eq.~(\ref{eq:dPdt}) through the sinusoidal factor in
the integrand. In the golden rule (\ref{eq:equiv}) the same factor 
is replaced with a delta function, but the figure shows that the 
effect of this replacement is significant and can not be neglected.

\section{Conclusions}

In conclusion we have shown that the net momentum imparted to a
trapped condensate by light scattering can be calculated by solving
the time dependent Gross-Pitaevskii equation in a regime which is
significant for current experiments. In the linear response limit,
this quantity is directly connected with the dynamic structure factor
for which the local density approximation (LDA) turns out to be a
reliable approximation in the explored range of $q$'s.  
The connection between the dynamic structure
factor and the quantities measured in stimulated light scattering 
experiments in the diverse accessible regimes deserves further 
investigations (see also Ref.~\cite{bob}). The present work
shows that, within appropriate limits, a quantitative analysis of
experimental data is indeed possible. More accurate and systematic 
measurements, including the explicit dependence of the momentum
imparted to the condensate on the duration of the pulse, would 
allow for a better test of the theoretical predictions for the 
dynamic structure factor of trapped Bose-Einstein condensates.

\acknowledgments

This work is supported by the Ministero dell'Universit\`a e della
Ricerca Scientifica. F.D. would like to thank the Dipartimento di 
Fisica di Trento for the hospitality.


\begin{center}
\begin{figure}
\input{pz-fig2.tex}
\caption
{The quantity $(4/V^2) P_z$, in units of $[\hbar \omega_\perp^2
a_\perp ]^{-1}$ is plotted against the frequency difference of the
two laser beams, $\omega$ in units of $\omega_\perp$, for  $q=3 
a_\perp^{-1}$ and two different
values of the time duration of the Bragg pulse $\tau_{B}$, in
units of $\nu_\perp^{-1} = 2\pi /\omega_\perp$. Solid lines 
are the predictions of LDA, as in Fig.~1. Dashed lines are the
results of the numerical integration of the GP equation (\ref{eq:GP}),
with Bragg intensity $V=0.2 \hbar \omega_\perp$. }
\label{fig2}
\end{figure}
\end{center}
\begin{center}
\begin{figure}
\input{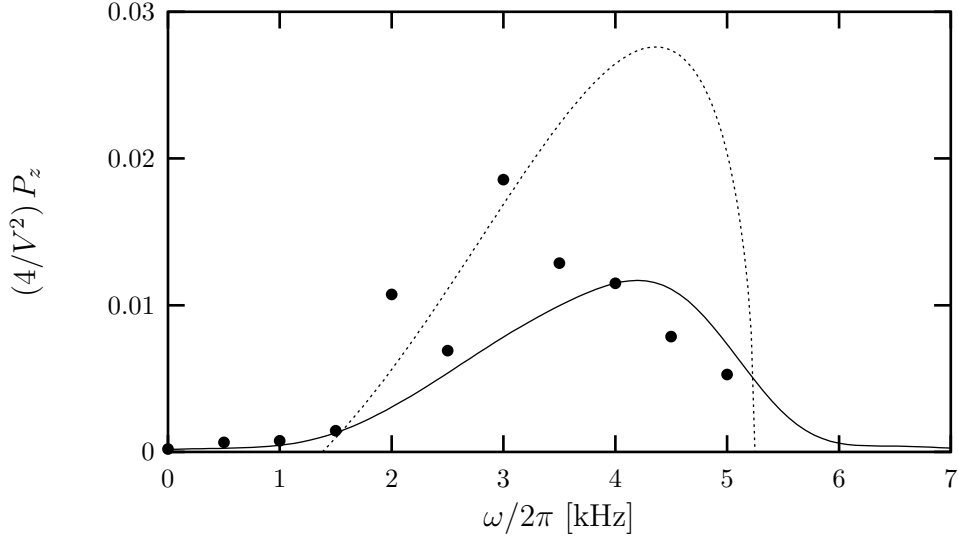}
\caption
{The quantity $(4/V^2) P_z$, in units of $[\hbar \omega_\perp^2
a_\perp ]^{-1}$ is plotted against frequency, $\omega/2\pi$, for $q=4.3
a_\perp^{-1}$ and $\tau_{B} = 0.06 (2\pi / \omega_\perp ) = 400\;\mu$sec. 
The solid line is the prediction of LDA, obtained from Eqs.~(\ref{eq:dPdt})
and (\ref{eq:SLDATF}) using the parameters of Ref.~\protect\cite{MIT2}. The
dotted line is the prediction of the golden rule (\ref{eq:equiv}). The 
experimental values of Ref.~\protect\cite{MIT2} (filled circles) are 
given in arbitrary units. 
} 
\label{fig3}
\end{figure}
\end{center}

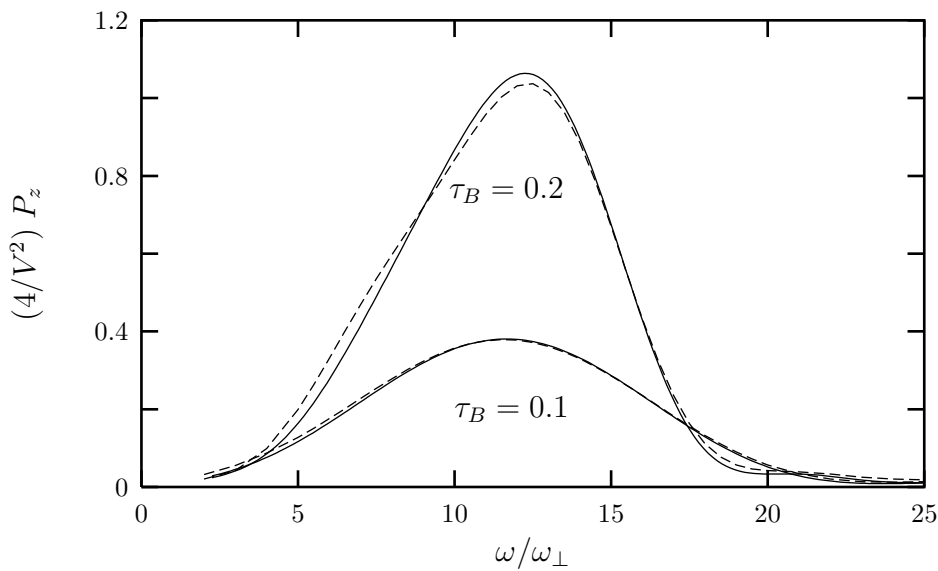
\end{document}